\documentclass[11pt]{article}
\usepackage[dvips]{graphicx}
\usepackage{amsfonts}
\usepackage{amssymb}

\title{Comparison of SDL and LMC measures of complexity: Atoms as a testbed.}
\author{C.~P.~Panos, K.~Ch.~Chatzisavvas\footnote{\texttt{e-mail:\, kchatz\,@\,auth.gr}}\,,
        Ch.~C.~Moustakidis,
        and E.~G.~Kyrkou
\\
 {\it  Department of Theoretical Physics,}\\
        {\it Aristotle University of Thessaloniki,}\\
                {\it  54124 Thessaloniki, Greece}
 }

\date{October, 2006}

\begin{document}

\maketitle

\begin{abstract}
The \emph{simple} measure of complexity $\Gamma_{\alpha,\beta}$ of
Shiner, Davison and Landsberg (SDL) and the \emph{statistical} one
$C$, according to L$\acute{\textrm{o}}$pez-Ruiz, Mancini and
Calbet (LMC), are compared in atoms as functions of the atomic
number $Z$. Shell effects i.e. local minima at the closed shells
atoms are observed, as well as certain qualitative trends of
$\Gamma_{\alpha,\beta}(Z)$ and $C(Z)$. If we impose the condition
that $\Gamma$ and $C$ behave similarly as functions of $Z$, then
we can conclude that complexity increases with $Z$ and for atoms
the strength of disorder is $\alpha \simeq 0$ and order is $\beta
\simeq 4$.
\end{abstract}

\section{Introduction}
There are various measures of complexity in the literature. A
quantitative measure of complexity is useful to estimate the
ability of a variety of physical or biological systems for
organization. According to \cite{Goldenfeld99} a complex world is
interesting because it is highly structured. Some of the proposed
measures of complexity are difficult to compute, although they are
intuitively attractive, e.g. the algorithmic complexity
\cite{Kolmogorov65,Chaitin66} defined as the length of the
shortest possible program necessary to reproduce a given object.
The fact that a given program is indeed the shortest one, is hard
to prove. In contrast, there is a class of definitions of
complexity, which can be calculated easily i.e. the \emph{simple}
measure of complexity $\Gamma_{\alpha,\beta}$ according to Shiner,
Davison, Landsberg (SDL) \cite{Shiner99}, and the
\emph{statistical} measure of complexity $C$, defined by
L$\acute{\textrm{o}}$pez-Ruiz, Mancini, Calbet (LMC)
\cite{Lopez95,Catalan02,Sanchez05}.

Whereas the perfect measure of complexity is as yet unknown, the
present work reports analysis of electron densities at atoms. We
specially refer to the shell structure (periodicity), using two of
the simplest of complexity measures, SDL and LMC calculated as
functions of the atomic number $Z$. This is a continuation of our
previous work \cite{Chatzisavvas05}. Those measures have been
criticized in the literature
\cite{Crutchfield00,Feldman98,Stoop05} and a discussion is
presented in Section \ref{sec:sec4}.

Our calculations are facilitated by our previous experience and
results for the information entropy in various quantum systems
(nuclei, atoms, atomic clusters and correlated atoms in a
trap-bosons) \cite{Chatzisavvas05},
\cite{Panos97,Massen98,Massen01,Lalazissis98,Moustakidis01,Panos01,Panos01b,Massen03,Moustakidis03,Massen05}.
A remarkable result is the universal property for the information
entropy $S=a+b\,\ln{N}$ where $N$ is the number of particles of
the quantum system and $a$, $b$ are constants dependent on the
system under consideration \cite{Massen98}. In fact, if one has a
physical model yielding probabilities which describe a system,
then one can use them to find $S$ and consequently calculate the
complexity of the system (in our case the atom) as function of
$Z$. This was done in \cite{Chatzisavvas05}, where we calculated
the Shannon information entropies in position-space ($S_r$) and
momentum-space ($S_k$) and their sum $S=S_r+S_k$ as functions of
the atomic number $Z$ ($2\leq Z \leq 54$) in atoms.
Roothaan-Hartree-Fock electron wave functions (RHF), for $1<Z\leq
54$, were employed \cite{Bunge93}. For $Z=1$ there is no
electron-electron effect. Higher values of $Z$ ($Z>54$), due to
the relativistic effects, are not considered. Analytic
wavefunctions are available in \cite{Bunge93} only for $Z$ up to
54. Their importance lays in the fact that we can derive accurate
wavefunctions in momentum space; in order to assure that the
integration is accurate. In \cite{Chatzisavvas05} we calculated
the SDL measure with RHF densities for $1\leq Z \leq 54$. In the
present work we calculate LMC measure in the same region of $Z$,
for the sake of comparison. One could consider the case of
Hartree-Fock wavefunctions for the extended region $1<Z\leq 102$
\cite{Koga02}. It turns out that both cases are satisfactory (work
in progress).

In \cite{Chatzisavvas05} and the present work, we examine if
complexity of an atom is an increasing or decreasing function of
$Z$. This is related with the question whether physical or
biological systems are able to organize themselves, without the
intervention of an external factor, which is a hot subject in the
community of scientists interested in complexity.

However, special attention should be paid with respect to the
meaning of complexity or structure, which may depend on the system
under consideration. In the case of atoms, the overall electrons
through pair-wise electron-electron interaction under the external
nuclear potential, lead to the characteristic electron probability
distribution. The selected information measure which shows up in
this manner is employed to estimate complexity. Periodicity is
clearly revealed here.

In Section \ref{sec:sec2} we review briefly measures of
information content and complexity. Section \ref{sec:sec3}
contains our numerical results for the LMC measure and discussion.
In Section \ref{sec:sec4} we comment on the validity of SDL and
LMC measures.

\section{Measures of information content and complexity of a
system}\label{sec:sec2}

 The class of measures of complexity considered in the
present work have two main features i.e. they are easily
computable and are based on previous knowledge of information
entropy $S$.

The Shannon information entropy in position space $S_r$ is
\begin{equation}\label{eq:equ1}
    S_r=-\int
    \rho(\textbf{r})\,\ln{\rho(\textbf{r})}\,d\textbf{r},
\end{equation}
where $\rho(\textbf{r})$ is the electron density distribution
normalized to one. The corresponding information entropy $S_k$ in
momentum space is defined as
\begin{equation}\label{eq:equ2}
    S_k=-\int n(\textbf{k})\,\ln{n(\textbf{k})}\,d\textbf{k},
\end{equation}
where $n(\textbf{k})$ is the momentum density distribution
normalized to one.

The total information entropy is
\begin{equation}\label{eq:equ3}
    S=S_r+S_k,
\end{equation}
and it is invariant to uniform scaling of coordinates, i.e. does
not depend on the units used to measure $\textbf{r}$ and
$\textbf{k}$, while the individual $S_{r}$ and $S_{k}$ do depend
\cite{Massen98}.

$S$ represents the information content of the quantum system (in
bits if the base of the logarithm is 2 or nats if the logarithm is
natural). For a discrete probability distribution $\{ p_i
\}_{i=1,2,\ldots,k}$, one defines instead of $S$, the
corresponding quantities $H$ and $H_{{\rm max}}$
\begin{equation}
 H=-\sum_{i=1}^{k} p_i\,\ln{p_i} \qquad \left(\sum_i p_i=1\right)
\end{equation}
and
\begin{equation}
H_{{\rm max}}=\log{k}.
\end{equation}
The uniform (equiprobable) probability distribution
$p_1=p_2=\ldots=p_k=\frac{1}{k}$, gives the maximum entropy of the
system. It is noted that the value of $H_{{\rm max}}$ can be
lowered if there is a constraint on the probabilities \{$p_{i}$\}.

Another measure of the information content of a quantum system is
the concept of information energy $E$ defined by Onicescu
\cite{Onicescu66}, who tried to define a finer measure of
dispersion distribution than that of Shannon information entropy.
Onicescu's measure is discussed in \cite{Chatzisavvas05}.

For a discrete probability distribution ($p_1,p_2,\ldots,p_k$),
$E$ is defined as
\begin{equation}
 E=\sum_{i=1}^k p_i^2,
\end{equation}
while for a continuous one $\rho(x)$ is defined by
\begin{equation}\label{eq:oni-1}
 E=\int \rho^2(x)\, dx.
\end{equation}
One can define a measure for information content analogous to
Shannon's $S$ by the relation
\begin{equation}
 O=\frac{1}{E}.
\end{equation}
For  three dimensional spherically symmetric density distributions
$\rho(\textbf{r})$ and $n(\textbf{k})$, in position- and
momentum-spaces respectively,  one has
\begin{equation}\label{eq:Er}
    E_r=\int_0^\infty \rho^2(r)\,4\pi r^2\,dr,
\end{equation}
\begin{equation}\label{eq:Ek}
    E_k=\int_0^\infty n^2(k)\,4\pi k^2\,dk.
\end{equation}

The product $E_{r}\cdot E_{k}$ is dimensionless and can be
considered as a measure of dispersion or concentration of a
quantum system. $S$ and $E$ are reciprocal. Thus we can redefine
$O$ as
\begin{equation}
    O=\frac{1}{E_r E_k},
\end{equation}
in order to be able to compare $S$ and $E$.

Landsberg \cite{Landsberg84} defined the order parameter $\Omega$
(or disorder $\Delta$) as
\begin{equation}\label{eq:omega}
    \Omega=1-\Delta=1-\frac{S}{S_{{\rm max}}},
\end{equation}
where $S$ is the information entropy (actual) of the system and
$S_{\textrm{max}}$ the maximum entropy accessible to the system.
It is noted that $\Omega=1$ corresponds to perfect order and
predictability while $\Omega=0$ means complete disorder and
randomness.

In \cite{Shiner99} a measure of complexity $\Gamma_{\alpha,\beta}$
was defined of the form
\begin{equation}\label{eq:gamma}
   \Gamma_{\alpha,\beta}=\Delta^{\alpha}\Omega^{\beta}=
   \Delta^{\alpha}(1-\Delta)^{\beta}=\Omega^{\beta}(1-\Omega)^{\alpha},
\end{equation}
which is called the ``simple complexity of \emph{disorder}
strength $\alpha$ and \emph{order} strength $\beta$''. One has a
measure of category I if $\beta=0$ and $\alpha>0$, where
complexity is an increasing function of \emph{disorder}, while
category II is when $\alpha>0$, $\beta>0$ and category III when
$\alpha=0$ and $\beta>0$ where complexity is an increasing
function of \emph{order}. In category II complexity vanishes at
zero \emph{order} and zero \emph{disorder} and has a maximum of
\begin{equation}
    (\Gamma_{\alpha,\beta})_{{\rm max}}=
    \frac{\alpha^{\alpha}\beta^{\beta}}{(\alpha+\beta)^{(\alpha+\beta)}}
    \quad \mbox{\textrm{at}}\,\,
    \Delta=\frac{\alpha}{\alpha+\beta} \,\,
    \mbox{\textrm{and}}\,\,
    \Omega=\frac{\beta}{\alpha+\beta}.
\end{equation}
Several cases for both $\alpha$ and $\beta$ non-negative are shown
in Fig. 2 of Ref. \cite{Shiner99}, where $\Gamma_{\alpha,\beta}$
is shown as a function of $\Delta$. In our previous work
\cite{Chatzisavvas05} we obtained $\Delta=S/S_{\rm max}$ or
$\Omega=1-\Delta$ as a function of $Z$ and plotted the dependence
of $\Gamma_{\alpha,\beta}$ on the atomic number $Z$.

We employed $S_{{\rm max}}$ according to rigorous inequalities
holding to atoms \cite{Gadre87,Chatzisavvas05}, which hold for
other systems as well (nuclei, atomic clusters and atoms in a
trap--bosons) as verified in \cite{Massen01,Massen02}. These
inequalities are

\begin{eqnarray}
S_{r\,{\rm min}}& \leq S_r & \leq S_{r\,{\rm max}}, \\
S_{k\,{\rm min}}& \leq S_k & \leq S_{k\,{\rm max}}, \\
S_{\rm min}& \leq S & \leq S_{\rm max}.
\end{eqnarray}

The lower and the upper limits can be written, for density
distributions normalized to one

\begin{eqnarray}
S_{r\,{\rm min}}& = & \frac{3}{2}\,(1+\ln{\pi})-\frac{3}{2}\,\ln{\left(\frac{4}{3}\,T\right)}, \nonumber \\
S_{r\,{\rm max}}& = &
\frac{3}{2}\,(1+\ln{\pi})+\frac{3}{2}\,\ln{\left(\frac{2}{3}\,\langle
r^2 \rangle \right)}, \\
\nonumber \\
S_{k\,{\rm min}}& = &
\frac{3}{2}\,(1+\ln{\pi})-\frac{3}{2}\,\ln{\left(\frac{2}{3}\,\langle
r^2 \rangle \right)}, \nonumber \\
S_{k\,{\rm max}}& = &
\frac{3}{2}\,(1+\ln{\pi})+\frac{3}{2}\,\ln{\left(\frac{4}{3}\,T\right)},
\\
\nonumber \\
S_{{\rm min}}& = & 3\,(1+\ln{\pi}), \nonumber \\
S_{{\rm max}}& = &
3\,(1+\ln{\pi})+\frac{3}{2}\,\ln{\left(\frac{8}{9}\,\langle r^2
\rangle \, T\right)} \label{eq:smax},
\end{eqnarray}
where $\langle r^2 \rangle $ is the mean square radius and $T$ is
the kinetic energy. We employ in (\ref{eq:omega}) $S_{\rm max}$
according to relation (\ref{eq:smax}).

As an alternative, we may use instead of $\Gamma_{\alpha,\beta}$
the following \emph{statistical} measure of complexity $C$ due to
L$\acute{\textrm{o}}$pez-Ruiz, Mancini and Calbet \cite{Lopez95}
defined as
\begin{equation}
    C=S\cdot D,
\end{equation}
where $S$ denotes the information content stored in the system (in
our case the information entropy sum $S=S_r+S_k$) and $D$ is the
disequilibrium of the system i.e. the distance from its actual
state to equilibrium \cite{Catalan02,Sanchez05} defined for a
discrete probability distribution $\{p_i\}$ as follows
\cite{Catalan02}
\begin{equation}
    D(\{p_i\})=\sum_{i=1}^{N} \left(p_i-\frac{1}{N}\right)^2,
    \quad \mbox{where}\,\, p_i\geq 0 \,\, \mbox{and} \,\,
    \sum_{i=1}^{N} p_i=1.
\end{equation}

$D$ is the quadratic distance of the actual probability
distribution $\{p_i\}$ to equiprobability. In the continuous case,
the rectangular function $\rho(x)=\frac{1}{2L}$, where $-L<x<L$,
is the natural extension of the equiprobability distribution of
the discrete case. Thus the disequilibrium could be defined as
\begin{equation}
    D^{*}=\int_{-L}^{L} \left(\rho(x)-\frac{1}{2L}\right)^2 \,dx=
    \int_{-L}^{L} \rho(x)^2 \,dx-\frac{1}{2L}.
\end{equation}
If we redefine $D$ omitting the constant adding term in $D^{*}$
(which is very small for large $L$), the disequilibrium reads now
\begin{equation}\label{eq:eq24}
    D(\rho(x))=\int_{-L}^{L} \rho(x)^2 \,dx,
\end{equation}
where $D$ is positive for every distribution and minimal for the
rectangular function which represents the equipartition. For large
values of $L$ ($L\rightarrow \infty$) relation (\ref{eq:eq24})
gives
\begin{equation}\label{eq:D-onedim}
    D=\int \rho^2(x)\,dx.
\end{equation}

Another derivation of the formula (\ref{eq:D-onedim}) for
continuous $\rho(x)$ can be found in Section 3 of \cite{Lopez03}
by using the R\'{e}nyi generalized entropy
\begin{equation}
    I_q=\frac{1}{1-q}\,\log\left({\sum_{i=1}^{N} p_i^q}\right),
\end{equation}
where $q$ is an index running over all the integer values.

According to \cite{Lopez95,Catalan02,Sanchez05} $S$ and $D$ are
the two basic ingredients for calculating complexity. In our
3-dimensional case, we employ instead of (\ref{eq:D-onedim}) the
formula
\begin{equation}\label{eq:D-3dim}
    D=E_{r}\cdot E_{k},
\end{equation}
where $E_{r}$, $E_{k}$ are defined in (\ref{eq:Er}),
(\ref{eq:Ek}). Relation (\ref{eq:D-3dim}) extends the definition
of measure of disequilibrium of the system according to LMC, to
our case, where we are interested jointly in position- and
momentum-spaces. It turns out that LMC definition of
disequilibrium $D$ (\ref{eq:D-onedim}) is identical to Onicescu's
formula (\ref{eq:oni-1}) for the information energy $E$. In fact,
inspired by the work of L$\acute{\textrm{o}}$pez-Ruiz, Mancini and
Calbet, a new interpretation of Onicescu information energy may be
proposed i.e. it represents the disequilibrium of the system or
distance from equilibrium. Additionally in our case something new
is introduced, that is the effect of a delicate balance between
conjugate spaces, reflected in the sum $S=S_{r}+S_{k}$ and the
product $D=E_{r}\cdot E_{k}$. Both $S$ and $D$ are dimensionless.
\section{Numerical results and discussion}\label{sec:sec3}

\begin{figure}[hb]
\centering
\includegraphics[height=6.4cm,width=7.4cm]{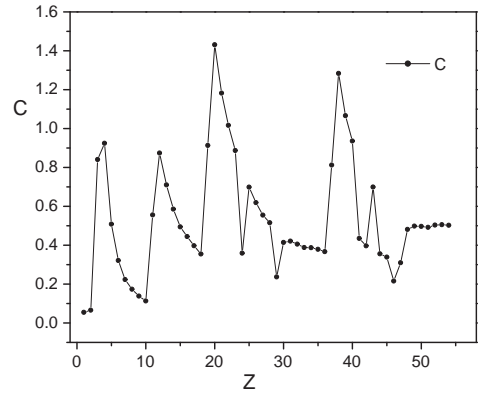}
\caption{The LMC measure of complexity $C$ as function of the
atomic number $Z$ of atoms.} \label{fig:1}
\end{figure}

\begin{figure}[ht]
\centering
\includegraphics[height=6.4cm,width=7.4cm]{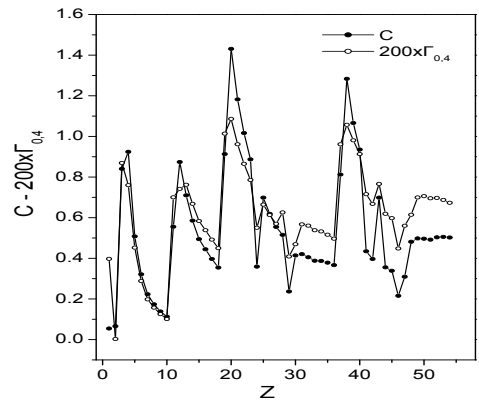}
\caption{Comparison of C(Z) and 200$\times\Gamma_{0,4}(Z)$}
\label{fig:2}
\end{figure}

The dependence of $\Gamma_{\alpha,\beta}$ on $Z$ for atoms has
been calculated recently in \cite{Chatzisavvas05}. In the present
letter we calculate $C(Z)$ employing the same RHF wave functions
in the same region $1<Z\leq 54$, for the sake of comparison. Our
results are shown in Fig. \ref{fig:1} and Fig. \ref{fig:2}. We
compare them with $\Gamma_{\alpha,\beta}(Z)$ shown in Fig. 3 of
\cite{Chatzisavvas05} for $(\alpha,\beta)=(1,1), (1,1/4), (1/4,0),
(0,4)$. In all (six) cases we observe that the measures of
complexity show local minima at closed shells atoms, namely for
$Z$=10 (Ne), 18 (Ar), 36 (Kr). The physical meaning of that
behavior is that the electron density for those atoms is the most
compact one compared to neighboring atoms. The local maxima can be
interpreted as being far from the most compact distribution (low
ionization potential systems). This does not contradict common
sense and satisfies our intuition. There are also local minima for
$Z$=24 (Cr), 29 (Cu), 42 (Mo). Those minima are due to a specific
change of the arrangement of electrons in shells. For example,
going from $Z$=24 (Cr) with electron configuration
[Ar]$4\textrm{s}^{1}3\textrm{d}^{5}$ to the next atom $Z$=25 (Mn),
with configuration [Ar]$4\textrm{s}^{2}3\textrm{d}^{5}$, it is
seen that one electron is added in an s-orbital (highest). The
situation is similar for $Z$=29 (Cu) and $Z$=42 (Mo). The local
minimum for $Z=46$ (Pd) is due to the fact that Pd has a
$4\textrm{d}^{10}$ electron configuration with extra stability of
electron density. It has no $5\textrm{s}$ electron, unlike the
neighboring atoms. There are also fluctuations of the complexity
measures within particular subshells. This behavior can be
understood in terms of screening effects within the subshell.  The
question naturally arises if the values of complexity correlate
with properties of atoms in the periodic table. An example is the
correlation of Onicescu information content $O$ with the
ionization potential (Fig. 4 of \cite{Chatzisavvas05}). A more
detailed/systematic study is needed, which is beyond the scope of
the present report.

Our calculations in \cite{Chatzisavvas05} show a dependence of
complexity on the indices of disorder $\alpha$ and order $\beta$.
In \cite{Chatzisavvas05}, we made a general comment that there are
fluctuations of complexity around an average value and atoms
cannot grow in complexity as $Z$ increases. The second part of our
comment needs to be modified. Various values of ($\alpha$,$\beta$)
lead to different trends of $\Gamma_{\alpha,\beta}(Z)$ i.e.
increasing, decreasing or approximately constant. In addition, in
the present Letter we compare $C(Z)$ with
$\Gamma_{\alpha,\beta}(Z)$ and we find a significant overall
similarity between the curves $\Gamma_{0,4}(Z)$ and $C(Z)$ by
plotting $C(Z)$ and $200\times\Gamma_{0,4}(Z)$ in the same Fig.
(\ref{fig:2}). The numerical values are different but a high
degree of similarity is obvious by simple inspection. There is
also the same succession of local maxima and minima at the same
values of $Z$. Less striking similarities are observed for other
values of ($\alpha,\beta$) as well, e.g. $\Gamma_{1,1}(Z)$ and
$C(Z)$.

Concluding, the behavior of SDL complexity depends on the values
of the parameters $\alpha$ and $\beta$. The statistical measure
LMC displays an increasing trend as $Z$ increases. An effort to
connect the aforementioned measures, implies that LMC measure
corresponds to SDL when the magnitude of disorder $\alpha\simeq 0$
and of order $\beta\simeq 4$. In other words, if one insists that
SDL and LMC behave similarly as functions of $Z$, then we can
conclude that complexity shows an overall increasing behavior with
$Z$. Their correlation gives for atoms the strength of disorder
$a\simeq 0$ and order $\beta\simeq 4$.

A final comment seems appropriate: An analytical comparison of the
similarity of $\Gamma_{\alpha,\beta}(Z)$ and $C(Z)$ is not
trivial. Combining equations (\ref{eq:omega}) and (\ref{eq:gamma})
we find for the SDL measure
\begin{equation}\label{eq:eq19}
    \Gamma_{\alpha,\beta}(Z)=\left(\frac{S}{S_{{\rm max}}}\right)^{\alpha}
    \left( 1-\frac{S}{S_{{\rm max}}} \right)^{\beta},
\end{equation}
while for the LMC one has
\begin{equation}
    C=S\cdot \left(E_{r}E_{k}\right).
\end{equation}
$S$ and $S_{{\rm max}}$ depend on $Z$ as follows
\begin{equation}
    S(Z)=S_{r}(Z)+S_{k}(Z)=6.257+1.069\,\ln{Z}
\end{equation}
(almost exact fitted expression) \cite{Gadre85,Chatzisavvas05},
while
\begin{equation}
    S_{{\rm max}}(Z)=S_{r\,{\rm max}}(Z)+S_{k\,{\rm max}}(Z)=
    7.335+1.658\,\ln{Z}
\end{equation}
(a rough approximation). We mention that $S_{r}$, $S_{k}$,
$E_{r}$, $E_{k}$ are known but different functionals of
$\rho(\textbf{r})$ and $n(\textbf{k})$ according to relations
(\ref{eq:equ1}), (\ref{eq:equ2}) and (\ref{eq:Er}), (\ref{eq:Ek})
respectively. It is noted that our numerical calculations were
carried out with exact values of $S_{r}$, $S_{k}$, $E_{r}$,
$E_{k}$, while our fitted expressions for $S(Z)$,
$S_{\textrm{max}}(Z)$ are presented in order to help the reader to
appreciate approximately the trend of $\Gamma_{\alpha,\beta}(Z)$
and $C(Z)$.

\section{Comments on the validity of SDL and LMC measures}\label{sec:sec4}

Complexity is a multi-faceted and context dependent concept,
difficult to be quantified. It is extremely difficult at present
to define complexity under all possible circumstances. Instead,
one can choose a pragmatic approach and use as a starting point
definitions of \emph{complexity} found in the literature and
attempt to evolve the existing framework. Thus, we have chosen the
\emph{simple} SDL measure and the \emph{statistical} LMC one,
which are relatively easy to compute. The question naturally
arises if the above measures do represent the concept of
complexity as expected from semantics, intuition or other general
criteria. SDL and LMC measures have also been criticized in
several aspects \cite{Crutchfield00,Feldman98,Stoop05}, some of
them are the following:
\begin{enumerate}
\item All systems with the same disorder $\Delta$ have the same
$\Gamma_{\alpha,\beta}$ i.e. the function
$\Gamma_{\alpha,\beta}(\Delta)$ is universal.

\item Landsberg's definition of disorder $\Delta=S/S_{\rm max}$
does not describe properly and does not capture the system's
structure, pattern, organization, or symmetries.

\item The calculation in \cite{Shiner99} of $\Gamma_{11}$ for
equilibrium Ising systems is not clear.
\end{enumerate}
In our previous and present work we use ground state electronic
densities of atoms to evaluate $\Gamma_{\alpha,\beta}(Z)$ and
$C(Z)$. At present it is not possible to answer semantic or
sophisticated questions as the ones raised in
\cite{Crutchfield00,Feldman98,Stoop05} and described above. One
cannot claim that SDL measure is \emph{the measure of complexity}
but it is a \emph{simple} measure, indicating that it can be used
as a starting point. To our knowledge, there are no other studies
of \emph{complexity} in quantum systems as functions of the number
of particles.

The definition of $\Gamma_{\alpha,\beta}$ is based on Landsberg's
disorder $\Delta$ in terms of information measures $S$, $S_{\rm
max}$ available as functions of $N$. By definition $\Omega$ and
$\Delta$ are related by $\Omega+\Delta=1$. An alternative approach
where $\Delta$ and $\Omega$ should be considered independent of
each other, is an open one. Before considering such an extension,
we calculate $\Gamma_{\alpha,\beta}(N)$ given in (\ref{eq:eq19}),
where $S$ and $S_{\rm max}$ show different non-trivial dependence
on $N$. In this sense the universality of
$\Gamma_{\alpha,\beta}(\Delta)$ is modified:
$\Gamma_{\alpha,\beta}(N)$ depends on the quantum system under
consideration, i.e. atoms \cite{Chatzisavvas05}, nuclei, atomic
clusters e.t.c. (work in progress). Instead of varying the number
of particles $N$, one could keep it constant, and study the effect
on \emph{complexity} of other parameters of the system. A welcome
property of a definition of complexity supporting its validity
might be the following: If one complicates the system by varying
some of its parameters, and this leads to an increase of the
adopted measure of \emph{complexity}, then one could argue that
this measure describes the complexity of the system properly.

There have been several criticisms regarding the LMC measure of
complexity in the sense that it does not deserve the adjective
\emph{statistical}, it is not a general measure that quantifies
structure and it is not an extensive quantity
\cite{Crutchfield00}. Analogous arguments with SDL could be given
to support the use of LMC as a first working prototype of
complexity to be modified by more sophisticated models. In
\cite{Stoop05} various properties of the SDL measure are examined
and a modification is attempted to remove its insensitivity to
system differences. However in our present (quantum) case of
atoms, by employing solely electron density distributions, it is
difficult to check these ideas.

In addition, the similarity of the qualitative behavior of
$\Gamma_{\alpha,\beta}(Z)$ and $C(Z)$ for atoms used as a case
study (although they obey different definitions) is interesting.
It shows that both measures share some common traits and correlate
strongly with the periodicity of the elements in the periodic
table. An important issue is the question if $\Gamma$ and $C$ are
true measures of complexity (structure, pattern etc) or are just
functionals of electron densities. Further research is needed to
clarify their region of validity and applicability as measures of
complexity.

Information entropy is an extensive quantity. The question arises
if $\Gamma_{\alpha,\beta}$ and $C$ are extensive or intensive
quantities, what is their thermodynamic limit, their behavior at
the boundaries (large or small order), their properties under
scaling transformations e.t.c. For example, it is stated in
\cite{Feldman98} that $C$ is neither an intensive nor an extensive
thermodynamic variable and it vanishes exponentially in the
thermodynamic limit, for all one-dimensional, finite-range
systems. However, in a broader context, $\Gamma_{\alpha,\beta}$
and $C$ can serve as indicators of complexity, based on a
probabilistic description. In some cases it can be applied in a
setting wider or different than thermodynamics. Such a case, is
the calculation of $C$ for various shapes of probability
distributions $\rho(x)$ in \cite{Catalan02}, i.e. the rectangular,
the isosceles--triangle, the Gaussian and the exponential
probability distributions which are classified according to the
corresponding values of $C$.

Another case is the present work, where $\Gamma_{\alpha,\beta}$
and $C$ are estimated for the ground state electron densities,
$\rho(r)$ and $n(k)$, of a quantum system (atom) at zero
temperature. To our knowledge, such an application is done for the
first time for a quantum system, and deserves more detailed
investigation (e.g. for excited states). It is true that there are
certain shortcomings of $\Gamma_{\alpha,\beta}$ and $C$ for
various cases in statistical mechanics. In quantum systems further
investigation is needed. Our title "\emph{$\ldots$Atoms as a
testbed}" describes well our aim.

\section{Acknowledgments}
The work of K.~Ch.~Chatzisavvas was supported by Herakleitos
Research Scholarships (21866) of $\textrm{E}\Pi\textrm{EAEK}$ and
the European Union. The work of Ch.~C.~Moustakidis was supported
by Pythagoras II Research project (80861) of
$\textrm{E}\Pi\textrm{EAEK}$ and the European Union.


\clearpage
\newpage


\end{document}